\documentclass[a4paper,11pt]{article}
\usepackage[utf8x]{inputenc}
\usepackage{fullpage}
\usepackage{amsmath}
\usepackage{amssymb}
\usepackage[dvips]{graphics}
\usepackage{graphics, epsfig}
\usepackage{amsfonts}
\usepackage{latexsym,amsmath,amssymb,graphics,amscd}
\usepackage{graphics,color}
\usepackage{graphicx}
\usepackage{boxedminipage}
\usepackage[toc,page]{appendix}
\usepackage{bm}
\usepackage{amsthm}
\usepackage{enumerate}
\usepackage{multirow}
\usepackage[table]{xcolor}
\usepackage{booktabs}
\usepackage{pict2e}
\usepackage{hyperref}
\usepackage{caption}
\usepackage{subcaption}
\usepackage{mathptmx}

\title{On Human Consciousness}
\author{Peter Grindrod  \\University of Oxford}

\begin{document}

\maketitle


{\it A thing of beauty is a joy forever}  {--- John Keats (1795--1821)}

\section{Introduction}
Human consciousness presents a number of particularly baffling challenges to  mathematics. On the one hand we know both the scale and the order of magnitude at which activity takes place in the human brain: there are billions on neurones, with each an active (excitable and refractory) electrical device, all coupled together within a characteristic network. The building blocks operate on the scale of cells, and  their synaptic interactions. On the other hand it appears to be difficult to describe what kind of information processing the whole might achieve, and also how this furnishes human beings with functionality usually associated with consciousness -- including the ability of the organism to perceive itself taking actions, or have feelings, or to envisage how it might act and reason within hypothetical circumstances: indeed the ability to create hypotheses beyond its present perceived reality - from real stimuli -  appears remarkable. 

The  theory of mind  literature  centres on defining aspects of what consciousness might be. It  describes many such aspects,  including the various properties of external perception, internal perception, intentionality and the existence of free will. It says rather little about how these  processes may actually occur, or their physical instantiation;  or why these phenomena are so robust and reliable within the human species. How hard can it be, since they are so universally achieved?

Emergence, or emergent behaviour, within coupled networks of (similar) interacting dynamical units (so called {\it complex systems}) refers to the appearance of macroscopic (higher domain) behavioural properties across such relatively large systems, often through robust phase change phenomena (as a function of scale or unit parameters, representing,say, operating conditions), that are often at first sight surprising or unfathomable from the perspective of an a consideration of component elements in isolation (the lower domain). Goldstein \cite{Gold} defines emergence as ``the arising of novel and coherent structures, patterns and properties during the process of self-organization in complex systems''. Philosophers generally term this {\it weak emergence}, since, though possibly {\it unexpected}, such properties may be demonstrated or deduced by very many, or some very complex, simulations and calculations, or analysis. Philosophers reserve the term {\it strong  emergence} for properties that are beyond any such deduction (even in principle) based on knowledge of the low level domain: in mathematical terms such phenomena must lie beyond any modelling and analysis. Most mathematical modellers would severely doubt if not reject outright that {\it any} such phenomena of real world systems exist, but would rather assert that, in any particular instance, it is the present modelling paradigm (from concepts to models to equations to analysis) that  fails to deduce the emergent properties, and thus it is not complete. Chalmers \cite{cha} thinks the opposite, and asserts ``there is exactly one clear case of a strongly emergent phenomenon, and that is the phenomenon of consciousness'' and says that `` facts about consciousness are not deducible from any number of physical facts.'' In contrast, the aim of this paper is to examine the extent to which s consciousness (at the high level domain of networks of neurones) are emergent and  already deducible from modelling considerations (at the low level domain of strongly connected directed networks of interacting neurone - see section \ref{sc}) and analysis. 


There is though a rather widespread objection amongst philosophers to the genesis of consciousness as an  emergent phenomena because without further any explanation it offers only a black box.  For scientific modellers such a discussion of real information processing and consciousness is akin to watching a man try to eat a steak without a knife and fork: it must be swallowed whole or else left alone on the plate. 

So to begin we shall  introduce some suitable  mathematical cutlery  with which to grapple with the challenge, including some relevant concepts from directed networks and dynamical systems that unlock behavioural functionality, that allow us to make some consequent predictions about the architecture that should drive efficiency at low cost. We shall discuss some functional properties of relatively small systems that form the basic layers within the  brain architecture, that has evolved specifically for real time information processing and  can underpin efficiencies in  both perception (and hence reaction) as well as some aspects of consciousness. 

We will suggest these elements impinge upon  considerations as to  how and why consciousness occurs, and what it is, which, though accepted as a necessary part of the consciousness conundrum by philosophers, is the major conceptual bottleneck. We are firmly within the {\it weak emergence} camp at this first stage, and we suggest  that the possibilities (some of which are discussed below) that conscious phenomena might be deduced from knowledge of physical and dynamical  structures are very far from exhausted.


The information processing architecture needs also to support aspects of memory and information retrieval that mean that newly perceived  events (in terms of new stimuli) can be related to known, prior, relevant  context and expectations, and thus interpreted and responded to very rapidly on the basis of only partial information. Thus the architecture must support a subconscious {\it best guess} as to context, reducing effort required and enabling rapid decision making at a very basic level, and a cascade of higher interpretations - what is happening, and how does it make us feel? - up to any reactive (automatic) or thoughtful (internal narrative-driven) response. 

In \cite{cha2} Chalmers makes a clear discrimination between the {\it easy problems of consciousness} and the {\it hard problem of consciousness}. In the early  part in this paper we are nibbling away at some of the easy problems, perhaps in a way anticipated by Chalmers. These are explanations for attributes of consciousness that can, or ought to, be explained scientifically. 

The hard problem is that of {\it experience}. What do we experience when we see and perceive things, actions, narratives? What is it like to be an organism that is perceiving the world? In section \ref{hard} we will address this issue directly and argue that the brain has a  {\it dual hierarchy} of perception. A dual hierarchy takes care of both the physical (external) elements that are perceived as well as the mental (internal) elements that are experienced.  On the physical side we have layers of elements: basic physical objects (sources of stimuli), to actions/events (relations between, or classifications of, objects), to narratives (relations between, or classifications of,  actions), to scenarios (relations between, or classifications of, narratives), and so on. The same exact discrimination and recognition process drives perception recursively at each level from the one immediately below, and outputs in terms of (re)actions may occur at each level. On the mental side we shall describe how internal elements must lie across the physical hierarchy and hence exist within a  {\it dual} set or {\it dual} space.

In section \ref{hard}, the central point of this paper,   we shall set out the conceptual structure (of the dual hierarchy model) in more detail, assuming the functionality available from  the earlier responses to the easy problem. We argue that the  dual hierarchies (physical, external,  elements and mental, internal, elements) are potentially infinite sets - there is always more to {\it know} and more to {\it experience}. And that each individual is on its  own a dynamical learning (and, sometimes, unlearning) curve. For any organism instantiating such a model we shall argue that there cannot be a well-defined mental experience (a feeling) that could not be represented by our mental elements; and in principle, with sufficient effort and exposure, be perceived.  This is crucial, for otherwise there would be at least some experiences that would rely on {\it other things} (things other than the  physical and mental element we describe in our dual hierarchy). What might these be? The only candidates are  firing patterns formed by spontaneous dynamical instabilities or else noise:  yet these could never be conjured at will by an organism itself, being subject to random chance and initiating perturbations. 

If  the organism is continuously learning and experiencing,  it has the potential to become more and more conscious, and to reach higher domains. In this way we see that consciousness is  an evolving, expanding, process rather than a very specific  state or benchmark to be achieved. Experiences possessing  only a finite number of facets could indeed be learned by zombies, so it is those experiences that give rise to a countable infinity of facets that could never be fully learned, even in a lifetime. 

We will address directly any idea that the mental elements defined within our construction, with a possibly infinite number of facets and nuances,  could not fully represent subjective experiences. 

Finally we shall summarise some of the implications for the dual hierarchy model that is set out in this paper. We  ague that internal mental feelings are akin to {\it latent} variables, which are so prevalent in control, behavioural pattern making,  and optimisation systems. They confer an evolutionary advantage in making agile discussions (top down) as opposed to response to observation (bottom up) in the absence of compete information. We feel that the most important element of this is not to be perfect but  to show that such candidates, based on all of what we know of how information is processed dynamically  within the cortex, do indeed exist. And that provided we focus on well defined experiences that are repeatable, and may be re-conjured by the brain, we can rule out any causal elements that are subject to spontaneous and uncontrollable internal dynamical events. 

\section{Preliminaries}
\subsection{Architecture, neuro-dyanamics and non-binary processing}\label{sc}
In a recent paper  \cite{GrinLee} the authors examine the behaviour of  {\it tight} bundles of coupled neurones, each having an excitable refractory dynamic, and arranged in a strongly connected directed subgraphs (SCDSGs), representing direct neurone-to-neurone connections, each incurring some time delay that is long when compared to the duration of a single neurone firing spike. It was shown that if the size (the number of neurones) of such an SCDSG varies whilst the expected degree distribution of the coupling remains constant (so that more locally, on the inside, all of the SCDSG  look the same) then the number of degrees of freedom exhibited by the (kick started and then) free running, autonomous, dynamics of an SCDSG has a dimension that grows only as fast as the logarithm of the number of neurones involved. In free running it generically behaves like a winding map on a $k-$dimensional torus, and all this happily independent of any particular choice of the ionic depolarising dynamical system (such a Hodgkin-Huxley or FitzHugh-Nagumo \cite{PaW}), even a discrete excitable spiking and refractory process model will suffice. 

\begin{figure}[ht]
\centering
\includegraphics[width=0.55\textwidth]{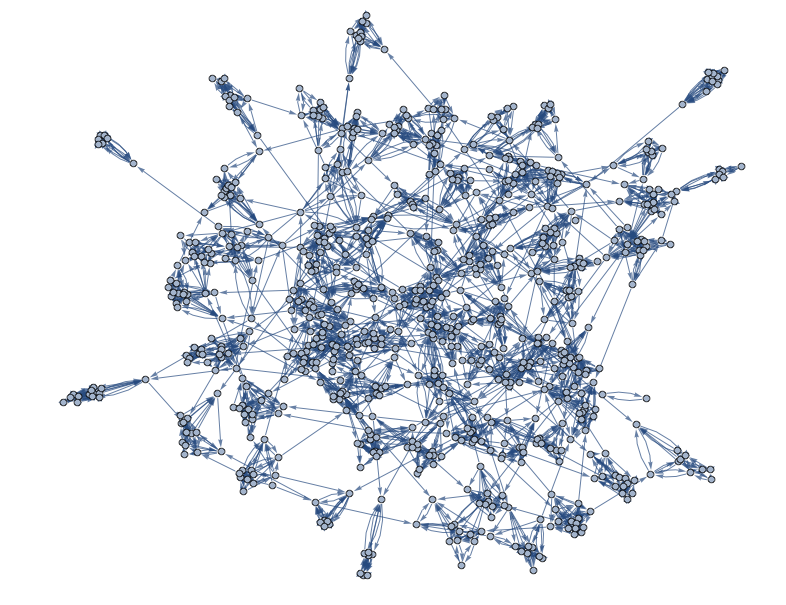}\includegraphics[width=0.44\textwidth]{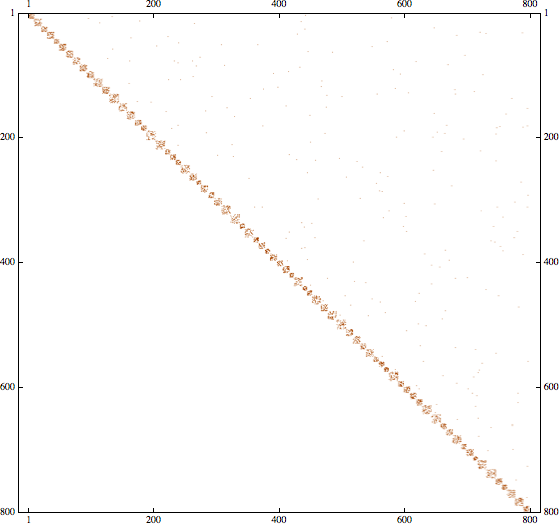}
\caption{A directed network containing many SCDSGs together with its block upper triangular adjacency matrix (from \cite{GrinLee}).}
\label{fig:SCG Network}
\end{figure}

The conclusion to be drawn is that within an evolutionary efficient brain we should observe many, many, such SCDSGs connected by a more loose network of macro, inter-SCDSG, connections. This situation is depicted in Figure \ref{fig:SCG Network} (taken from \cite{GrinLee}). Moreover, given obvious  volume and energy constraints we expect to observe brains having large numbers of smallish SCDSGs  rather than occasional SCDSGs of arbitrarily large sizes or even a giant component. Put more simply, each SCDSG may be stimulated to behave in time in a number of alternative modes, yet that number grows only sub-linearly with its size. The modes themselves are  dynamical patterns (across the network and over time)  and they are competitive one with another, They are not superposable and when the SCDSG is stimulated periodically (from another upstream SCDSG) the it may respond and become``resonant" with that  input (in a suitably selected mode) or else display more disordered behaviour. Thus the response of an SCDSG to upstream stimuli and which of its nodes might dominate depends where it is stimulated (what is the entry or receiving neurone) and the frequency of the stimulation. Only a single mode can respond, dampening or locking out the others, and when it does so only certain neurones may be heavily involved resulting in the routing of stimulation downstream to other SCDSGs that are immediately downstream of those participating neurones. This an SCDSG acts as an analog filter, a dynamical decision maker preferring one or another resonant mode, and amplifier and a router. That is how SCDSGs process incoming stimuli to produce outputs. Thus SCDSGs are the building blocks for an information processor. By definition this processing is non-binary, with multiple modes within each {\it processor} (SCDSG) competing on the weight and repetitivity of stimuli, just as hypotheses might compete within a Bayesian multiple hypotheses updating process.

To summarise, in \cite{GrinLee} it is proposed that an evolutionary efficient brain, maximising the range of processing behaviour available for given volume and energy constraints, wold be in the form of many loosely networked smallish SCDSGs, and that the information processing is no-binary with each SCDSG able to existing within a number $k>>2$ states. Moreover that the SCDSG can be reused to relay certain types of signals from distinct inputs through to corresponding outputs (for the right frequency of stimuli).   

Cascades and networks of winding dynamics over tori driving one another is a topic akin to that of coupled oscillatory systems where cascades of generalised clocks may drive one another in either type 0 or type 1 fashion \cite{ClocksToChaos}. Yet the generalisation to tori and the classification of different types of entrainment is relatively unexplored \cite{GrindPatel}.

For continuum models, such as generalisations of Amari models \cite{Amari},  the importance of transmission delays in coupling has been partly investigated and even in simplified and homogeneous circumstances give rise to exotic stability (spectral) behaviour, with multiple modes that collapse when the delays tend to zero \cite{GrinPin}.

We should perhaps also introduce a level of uncertainty (variability and temporal volatility) into the performance of both intra-neuronal cellular dynamics  as well as possible neurone-to-neurone transmission. This occurs not least due to the ionic distributions within  both extra-cellular and intra-cellular media. Clearly the conditions may fluctuate in time and across the networks (as a result local and whole brain performance may be influenced by the control of neurotransmitter distributions). Consequently the cells are not sitting within a constant environment: it is both heterogeneous and evolving (perhaps on timescales close to those of the neurone-to-neurone interactions). Thus many connections may fail or else become enhanced as local the ionic conditions change. Similarly the neurones themselves may be more or less ready to fire when stimulated. 

For this reason we may wish to consider some stochastic generalisations of the {\it central} deterministic dynamical systems introduced above. Already we have described a system that is complex (due to its size and the nature of its coupling) and is also multimodal in its outputs, even for a single SCDSG, as a function of any driving stimuli. Now we must also accept that some stochastic processes will be at work in any description at the time and spatial scales we have introduced here. However for the avoidance of doubt we do not propose to rely on such fluctuations in local performance as a means of introducing individualistic behaviour by some sleight of hand. We shall contend that without such perturbations the {\it central} dynamical systems introduced above is enough to allow the emergent of element equivalent to some aspects of consciousness.

Finally we note that an information processor built on a non-binary basic layer should be good at making decisions, yet probably  poor (inefficient) in both logic and arithmetic. Binary computers are the complete reverse. Of course in principle one might build one from another (as in deep learning) but not without requiring much capacity and much processing. The latter seems highly unreasonable and it should be horses for courses. Binary computers should do binary things; non binary computers should do non-binary things such as multi-hypothesis decision making based on partial information.

\subsection{Implications for large scale brain simulation and non-binary computing platforms}
The narrative analysis given in  \cite{GrinLee}  rests on the very large number of dynamical simulations of SCDSG of sizes varying by orders of magnitude and stimulated in different ways. The assumption that there are (real valued) transmission delays that all are large compared to single pulse events is crucial in forcing the degrees of freedom exhibited to be large. If delays are  merely integers, or else are all set to unity (as is sometimes the case in attempts to build arrays of linear or nonlinear non-binary processors), then the sophistication of the SCDSGs  
simply collapses. That observation itself  has clear implications for non-binary computing platforms (such as \cite{ DARPA}) and for large scale brain simulation programmes (such as the EU Framework 7 Human Brain Project \cite{THBP}). The heterogeneity of the neurone-to-neurone connections is far more important than one's particular choice of excitable-refractory neuronal dynamics. Ironically the solution of  large systems of delay differential equations (with a plethora or real valued delays) is very expensive (in terms of both compute time and memory) to achieve on a standard binary computing device: yet it is achieved electro-chemically when instantiated (hardwired by neurones) within the brain itself.

\subsection{Friston's {\it Bayesian Brain}}
The Bayesian brain is not a new idea and indeed many of the constructs developed in \cite{GrinLee}  were foreshadowed by ideas and discussions set out by Friston (in  \cite{Friston2013} and the references therein) coming from a completely different starting point. Instead of SCDSGs, Friston proposes active bundles of (implicitly strongly connected) neurones isolated within a Markov blanket. The dynamics of the active cells is expected to be  complex (usually illustrated with chaotic, chattering, attractors), whereas in \cite{GrinLee}, when left to free run, the SCDSGs result in winding dynamics over lowish dimension tori. The observation that the number of  degrees of freedom achievable within the dynamical behaviour does not scale with size (given that the directed networks must appear locally similar)   is very advantageous to, and supportive of,  Friston's conceptual  arguments and is also of practical importance (in bounding the local dimension of any observable  attractors) when examining EEG and other observable signals within neuro-imaging.

Friston goes much further explaining how such an architecture not only generates responses to common and  distinct incoming stimuli, like decision-making  at a very fundamental level (by selecting out ``modes'' in our discussion above), but also evolves the modes to do so ever more efficiently. This requires a feedback loop, and hence reward systems, that makes filtering and response more efficient (and less likely to wobble, slip,  or to tune out) by allowing multi-SCDSG (macro-level) resonance/cooperation to trigger the release of chemicals that enhance those presently activated connections within cooperating SCDSGs. We shall say nothing \label{Frist} about this tuning process here  except that it can build upon the architecture and permissive non-binary processing introduced above. We  shall though  consider the  mathematical consequences of such self-tuning, since not only does it  enable common feats of processing to become more efficient, it may also  enable a feat of self-recognition to take place, spontaneously.

\subsection{Physical structures}
The cerebral cortex is the brain's outer  layer of neural tissue in humans and other mammals, which wrapped over the limbic brain (which controls more basic functions). The cerebral cortex is usually asserted  to  play a role in memory,  perception,  thought, language, and consciousness.  

The human cerebral cortex is folded, increasing the surface area to volume ratio, with more than two-thirds of the cerebral cortex surface within the folds. The neocortex (the most recently developed part) is differentiated into six horizontal layers (with older parts being differentiated into fewer layers). Neurones from down within the different layers connect vertically to form  small {\it circuits}, called cortical columns. It is these tightly connected columns that are suggestive of a massive number of  SCDSGs,  introduced above, with each acting as a multimodal processor, yet integrated  more loosely across surface of the cortex. There is thus physical evidence for the architecture that is suggested to be efficient by the graph theory and predictions given in \cite{GrinLee}. 

\subsection{Quantum mind hypotheses}
It is relatively common in mathematical physics to engage with proponents of the quantum mind hypothesis in one form or another. This proposes that consciousness is beyond  classical approaches, that are based on dynamical systems describing a systems of coupled  networks of deterministic or stochastic dynamical systems (such as those introduced above), and that therefore quantum mechanical phenomena  must be at work.  In  \cite{Tarlaci} the authors discuss whether such quantum physics has significance for the nervous system. However hard calculations \cite{Seife},  combining data about the brain's temperature, the sizes of proposed quantum objects, and disturbances caused by ions, show that collapsing wave functions of tiny structures decohere  too rapidly to orchestrate the firing of neurones.

We agree with  Chalmers \cite{cha2} that ``The attractiveness of quantum theories of consciousness may stem from a law of minimisation of mystery: consciousness is mysterious, and quantum mechanics is mysterious,  so maybe the two mysteries have a common source.''

In this paper we shall adopt the view that the assumption of any quantum effect is simply not necessary to drive the emergence of functional  information processing, behaviour that allows a human brain to perceive itself (and to create hypotheses about itself). We suggest that there is simply no need to resort to such a hypothesis and moreover that the time scales of all neurone-to-neurone interaction and the passage and processing of information in the form of spiking behaviour around neuronal networks (dynamical or stochastic) preclude it. We assert that we have yet to  exhaust the possibilities of the large scale dynamical system approach.  

The argument that the human mind  is not running a computable algorithm, which goes back to Penrose \cite{pen},  ascribes a level of rigour (in terms machine-based computability) to rather approximate human subconscious decision making that is simply not warranted. This is evident from the evidence of nonoptimal decsion-making exhibited with behavioural economics, discussed in the next subsection.

\subsection{Subconscious heuristics }
The rise of behavioural analytics and the demonstration of the use of a range of heuristics as a means of making fast, possibly biased, and certainly suboptimal, subconscious decision making \cite{Kahn, Ariely} is ample demonstrated by the human brain's failure to reason (subconsciously) logically or quantitatively in even relatively simple circumstance. The huge success of this fast reasoning paradigm in explaining a very wide range of human cognitive biases and illusions, including loss aversion, yet hastening response by massively reducing  the cognitive effort required in any challenging circumstance is evident. Therefore we argue that the level of rigour (and any objective accuracy) involved in information processing is not high and the biased (and certainly suboptimal) nature of subconscious decision making evident within behavioural analytics scotches the idea that the mind can be subject to fine discriminations on the nature of computability.

The evidence of the success of heuristics in enabling fast, subconscious, decision making lends us to suggest that the most essential element that is required within any model of the brain's information processing is that multiple hypotheses must be competed based on both partial  and missing information  (the heuristics reflect this - they are observable consequences of it). The architecture and functionality discussed in \cite{GrinLee} indicates that this can happen at the scale of single SCDSGs, at the most basic level in the architecture. Moreover this competition is essentially a hardwired  approximation to Bayesian multiple hypothesis testing (somewhat inaccurate yet functionally similar), resulting in a single winning mode (the decision to adopt a hypothesis), and consequent information routing.

\section{What work can the machinery of cortex do?}
We have argued that each column (that is each SCDSG) is a low level decision maker, competing multiple hypotheses. When driven at some point of entry from up stream it either acts incoherently or else it locks into one of many modes. Consequently certain neurones within the column become relatively more involved and then information can propagate, through them, downstream on to other columns. Thus each column acts as a non-binary processing filter and consequently a  directional router for distinctive types of patterned behaviour.  What can a whole mesh of such objects achieve?

The most obvious response is that of searching for and distinguishing between different types of patterns. The human brain is remarkably good at this. And the same ability can be used for a range of distinct inputs. Thus brains should discriminate between visual patterns (just as we search for patterns in random images or when we look at clouds) as well as audio patterns within language and music. The latter are most interesting since they have been developed by humans to be learnable, pleasurable and evocative. We may think of language as the {\it negative} image of the brain's {\it positive} machinery. There are no languages that are difficult for human babies to learn, they have evolved structures which are intrinsically learnable and well suited to the brains machinery (the processing available within  the cortex). 

In fact in recent unpublished work \cite{Patel} we found ways to recursively  generate pairs of artificial grammars (over a binary alphabet) that  were indistinguishable based on their autocorrelation structures up to arbitrarily large orders (lags). Thus any signal processing methodology for discrimination would be costly in terms of memory alone (never mind how the computations might be achieved). Yet when tested individuals  could easily do better than random in distinguishing between these pairs of languages. Clearly the brain machinery unpicks the recursive definition in some way or else must alight on occasional signatory motifs that occurs in some but not other languages. This requires much more experiment. 

Under either option  the machinery must be able to make a decision that patterns are indeed embedded in the data that match stored, remembered patterns, observed earlier to which a label has been associated; or else the new data contains none of the previously known patterns, and thus could be discarded or else used to create a new remembered template if there is a strong recurrent pattern within it. Thus the cortex needs to hold a lexicon of {\it motif} patterns as certain firing modes with and across its columns, as well as being able to search out any one such pattern and to select that  most likely to be dominant and hence  associate the incoming stimuli with the mode. In doing so itself it might be best tactically to  analyses the incoming data recursively, at multiple resolutions,  both discarding what is inconclusive or commonplace and homing in on points of peculiarity/idiosyncrasy.

We suggest that the machinery might instantiate a (hard wired) methodology that tunes into a number of modes continuously  that is similar to the EM algorithm \cite{EM}: a permissive Bayesian-style updating competes and decides between a number of hypotheses (the E-step -- this is what was observed in \cite{GrinLee}) and then loosely and continuously re-calibrates (updates) the modes corresponding to each hypothesis (the M-step), accounting for the  most recently stimulus assigned to that particular mode. Of course the latter re-tuning step must rely on the reward process mentioned in section \ref{Frist}. 

The EM algorithm and its generalisation provides a solution to finite mixture modelling where the totality of all incoming signals are described in terms of a finite mixture of signal from a number of distinct sources (that in this case correspond to the distinct dynamical modes).

Such algorithm may be quite unwieldily to programme onto a standard digital computing platform: but we suggest that the cortex machinery (as a platform) may make this functionality relatively trivial, being only one level up from the basic M-step functionality available at the SCDSG level.

\section{The hard problem of consciousness}
What is it like to be an organism that is perceiving the world? \label{hard}Chalmers  \cite{cha2} suggests that this problem is (and may remain) beyond deduction, beyond any  scientific explanation. Here we shall introduce a conceptual model, building on the task-based  processing and multi-modal decision making properties of the brain's architecture. We will speculate that the successive sophistication of the {\it elements} that may be perceived, is based on a  {\it dual hierarchy}. 

Starting with a simple hierarchy, based on the successive integration and perceived understanding of external information  (and upon stored contexts and expectations) dealing with physical, objects, events, and higher domain structures, that are external to the brain. We shall set this up and then argue that internal elements (mental  objects, events, feelings and so on) must occupy a second hierarchy that is orthogonal to, or a dual of, the first hierarchy.  

\begin{figure}[ht]
\centering
\includegraphics[width=0.9\textwidth]{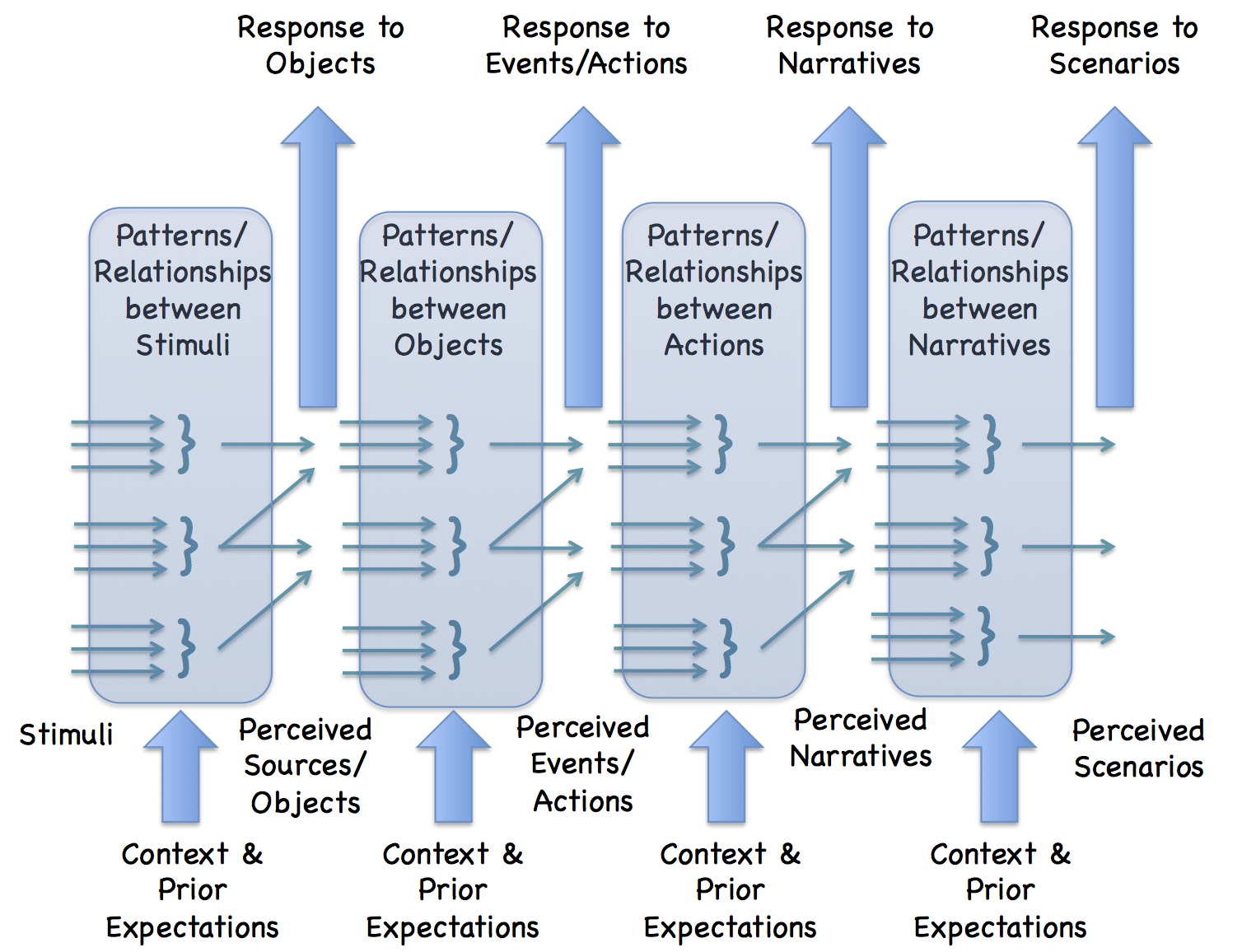}
\caption{Hierarchical arrangements of recursively derived real world (external) modal elements: from distinctinguished objects through to distinguished scenarios.}
\label{harp}
\end{figure}

Consider the perception of physical elements that might be taking place around us.  Each element perceived by the brain is an ``output'' pattern that wins a multiple hypothesis competition within a particular SCDSG, and these may be put into a separate of classes. The classes of elements might range from basic external objects (that are perceived to be the immediate sources of incoming stimuli), to actions/events (that is, relationships between, or classifications of, objects), to narratives (relationships between, or classifications of,  actions/events), to generalised scenarios (relationships between, or classifications of, narratives: responses to the perception of narratives), and so on. The classes make up successive layers within a hierarchy. This is illustrated in Figure \ref{harp}. We are not wedded to the number of layers shown, nor the nomenclature employed here.  In this case we show four levels:  objects, actions, narratives and scenarios. But the point is that the elements at each level are  defined recursively in terms of the various elements from the level immediately below, and drawing upon context and expectations (biases) defined from associated memories. Each element is a function of a multimodal analysis (decision making) of the previous layer.  The machinery that achieves this is identical between all successive levels -- and readily available, as we have pointed out. The recursion is very natural, robust and repeatable, using the circuitry we have discussed above;  yet it is rather hard for analysts to conceive of the consequences of this. 

The vertical boxes shown in Figure \ref{harp} represent the  successive pattern recognition tasks exploiting both remembered contextual information and prior expectations from past events (drawn  subconscious form memory -- in an associated network fashion) as well as the assumption of the structures (elements) that are identified at the previous level. Its output is the assumption of the structures (elements) identified at the current higher level. The previous output element from the lower levels are thus the dynamic inputs to the preset level. The machinery, a permissive type of  Bayesian discrimination and decision making (that we have established above, and that is readily available within the brain architecture via SCDSGs),  is the same in each case. The elements perceived become more sophisticated and more abstract or generalised as things move up the hierarchy (to the right in Figure \ref{harp}). 

For example, given our prior history in recognising certain objects in a present context (a church service say), some visual, auditory, and olfactory stimuli,  together with retrieved context and expectations, may  enable us to recognise the following objects: choirboy,  flame, hair, shouting, burning smell, candle. These objects together with further retrieved context and expectations allow us to recognise the following events/actions: candle burning, hair on fire, choirboy shouting. These actions together with further retrieved context and expectations allow us to recognise the following narrative: ``The burning candle has set the choirboy's hair  alight causing him to cry in pain and fear.'' This narrative together with further retrieved context and expectations allow us to perceive general scenarios: ``an accidental emergency'', ``choirboys always misbehaving'', ``an extreme event''...


There is no mystery within this process here: but the recursive nature  of the hierarchical process and the move from objective information (the perceived real world objects) through to subjective decisions  that are both more abstract and subjective, means that we find it successively more difficult  to be specific about the higher domain elements.

Now consider some simple internal (mental) elements: say, mental objects such as ``threat'', ``arousal'', ``excitement'', ``amusement''; or mental feelings such as the experience of the ``blueness of blue''. You cannot perceive these from any external stimuli: they are private and internal (but you might communicate them by your consequent behaviour). These mental elements do not fall into a single layer within the hierarchy introduced above for physical elements.  The perception of ``threat'' may be associated with physical objects (``flames" and ``burning"),  yet also with physical events (``hair setting on fire''),  with the particular physical narratives, and also the more generalised physical scenarios classifying the narratives (``an extreme event''). The mental object, ``threat'', may trigger immediate (physical response to danger - the release of adrenalin, for example) but it must also sit within its own hierarchy of classes (layers) of mental (internal) elements. 

Thus these must be within a second classification of elements that is orthogonal to the first one. Again we expect it must have multiple layers  representing  more and more abstract or conceptual mental elements, further away from simple element of mental consciousness. We  construct it as follows.

We  assert that set of all mental elements is isomorphic to the power set of the set of all physical elements: the set of all subsets of physical elements, drawn from across any and all layers within the physical hierarchy. Each subset of physical elements constitutes a potential mental element. Alternatively we are saying that any  mental element corresponds precisely to a particular subset of those physical elements (that may be drawn from across any and all layers within the physical hierarchy) that, in combination(s),  invoke or inspire it, within  real time.  As with the physical elements hierarchy, the mental elements  will also take inputs from stored contexts and expectations. These element are generalised feelings and experiences.  The characterisation of the set of all mental elements as the power set of the set of all physical elements is crucial step. Note the mental element hierarchy is of a much larger size, or of higher cardinality, than the set of physical elements, since the latter is not  isomorphic to its own power set.

We may if we wish induce a hierarchy for all such metal elements by placing each at  the mental level corresponding to the maximin physical level of any of its corresponding inspiring subset of real elements.  Thus the set of all potential mental elements  has a natural  hierarchy  which is induced from the the hierarchy over the physical elements, which was  defined recursively. 

We have argued that the power set of the set of physical elements is the appropriate structure, but others may wish to think of this a  dual set, or {\it dual space}: that is, the set of functionals that can be defined over the set of all physical elements. At this point the difference in these concepts seems far less important than the fact that suitable candidates for the definition of the set of mental elements indeed exist.

Unlike the physical elements hierarchy the mental elements hierarchy is not recursively defined though. Instead we have defined  mental elements  in terms of the subsets of the set of all physical elements that will (in combinations) invoke or inspire them. 

Now it may be agued that we are yet to  give any reason to think that the mental elements in the  mental hierarchy (the power set of all physical elements) should be associated with subjective experiences.  So let us summarise the reasoning  that this must  indeed be so. 

Why must the  mental elements,  defined as the power set of the set of physical elements, be equivalent to our subjective (internal) feelings? 
\begin{enumerate}[(i)]
\item We have argued that the feelings and experiences cannot always be placed within the physical elements hierarchy: though each is invoked or inspired by perhaps quite large set  of physical elements -- actions, events, scenarios, and so on; with some of these physical elements possibly at a high, abstract, level within the physical hierarchy. Therefore our feelings in general must exist across the physical events hierarchy, as required combinations of physical elements to be perceived so as to  inspire them. Similar combinations of real elements may well define similar feelings, so we may have  generalized feelings ({\it unease/dread} for example), representing classes of feelings - broad supersets, or sets of higher domain physical elements.
\item Why should sets (combinations)  of physical elements define subjective feelings? Every feeling that we have ever had has been in response to some original combination of external physical events: these (first time experience) physical elements together defined the feeling, within the moment. Indeed if we re-perceive, or remember,  the right set of  corresponding physical elements (a specific  event, a situation, an image, a fact, a narrative, ...), in the right combinations, then we can re-experience the desired mental feeling.  We can make ourselves feel sick with disgust, or shocked, or filled with awe. But we do not do so in the abstract (we do not will ourselves to feel something and flick a mental switch):  we must mentally re-conjure some combination of physical processes, by thinking of a real or hypothetical set of physical elements. With practice we might become very proficient at this and also possibly  hardened to certain experiences (reducing the initial physical reactions to them). Therefore our feelings are indeed invoked or inspired by certain sets of physical elements (even if they not all actually really present, but imagined). They are really just a classification of combinations of physical elements (some of which might be abstract and complex - relations between relations between relations).
\item Note that we may also need to be participating in some action or take care to be in the right state of mind to be able to invoke/inspire a feeling -- but these are again component elements, and ultimately actions provide us with sets of physical elements. 
\item Why must all feelings be inspired in this way, and be defined in terms of a combination of physical elements?   Put more simply: can we have a feeling that is not well-defined in this way? Can we have a feeling that requires anything else, additional? Sometimes feelings can be combinations of other feelings; but that does not count as an exception. 
Suppose now that there is a particular feeling that is contingent on something other than a set of physical elements (or mental feelings that are already defined in terms of physical elements). How did that experience come about inside our minds for the first time, what precipitated it? There must have been a set of circumstances and physical stimuli, from our own senses or the workings of our own body (as an organism), that communicated and instigated  some information, as patterns of neuronal behaviour,  into our brain. But this information {\it is} physical and it {\it defines} physical elements.  If there were anything else it would have to produce some spontaneous patterns of firing behaviour across some or many SCDSGs. As non-physical this pattern formation might be internally created, such as in instability-driven pattern formation, like Turing patterns, following spontaneous symmetry breaking within certain modes, or as a last resort,  noise (chatter between neurones). We cannot rule this out: but either of these would be uncontrolled aberrant firing patterns in time and across the cortex, created by random perturbations and instabilities; or  noise within the system. Yet these can not give rise to {\it feelings} as they plainly cannot be re-conjured at will. They are akin to functional breakdowns (aberrations) rather than facets of the organism's regular experiences. So let us restrict our definition of feelings to those subjective experiences that can be relived at will.  Then we must discount internal instability driven pattern generation and noise as causal element of feelings.
\end{enumerate}

Of course any individual may or may not have perceived any particular mental elements (as yet). They may or may not have perceived the requisite combination (subset) of the physical elements. So each organism will have its own version of the mental element hierarchy depending on the sophistication of its real experiences, as held within its current version of the physical element hierarchy. We must understand that both sets, that of perceived physical elements, and that of the  induced  mental elements,  are constantly growing with the experience and the familiarity of experiences or the organism.

There is no {\it extra ingredient} here:  the processing is applied recursively and is updated at all levels within the  real element hierarchy. Both hierarchies can be updated simultaneously. However the dual hierarchical model certainly helps us to understand how elements become more remote from the simple real elements (low domain real objects and events) and the simple mental elements  (induced by just a few  low domain real objects and/or  events),  and hence they become more subjective and abstracted, whilst also being as ``real'' to the brain as are much lower domain objects and events.


There are always  more layers in the hierarchies,  possibly a countable infinity of layers due to the reclusive definition in both hierarchies. As the layers of elements become more abstract and more subjective (and experiential) they may be many layers, which are harder and harder to discern one from another. 
It will almost  certainly be the case that there are higher layers in the metal hierarchy repressing super-experiences,  may emerge from  very distinct experiences  and specific exposures of the organism to different types of situations (the mental elements hierarchy might be  itself constantly developing and is not assumed to be hard wired with a fixed number of levels). Some of these super-experiences may be developed by individuals with specific exposure  and practice that is unusual for others (such as experts in perfume, taste, and music). It may be that certain claimed religious experiences gained via mediation or other activities are simply developing and accessing these higher and higher layers of experience within the hierarchy. Thus all similar organisms may have the potential to develop more and more super-experience (higher domain) layers within our proposed mental hierarchy: it is very likely that only the lower levels will be most common (up to and including common experiences, common qualia). Yet certain individuals, through their own effort or exposure, may develop such super-experience layers of processing that are accessible for them  as a consequence of  certain types of stimuli and internal (biased) retrieval. 

The high-level mental elements such as feelings  or experiences may feed-back into the control of the retrieval of contextual information and prior expectations that are drawn into the pattern recognition and multimodal design making earlier down within both  hierarchy. Thus the whole has a feedback loop: if we experience feelings of extacy,   or  rage then that will constrain and bias the kind of information we retrieve in clarifying the present context, thus recognised elements at low levels will be updated (dynamically) partly in the light of the perceived higher domain elements.

Notice that if set of the real  elements were to be finite, once and for all, say $n$  of them, then so would the set of mental elements: there would be $2^n$ of them. Then a robot or zombie could learn all of them by rote. But the physical hierarchy  is defined recursively so that each layer can be generated as required from the lower domains  given the right combination of stimuli (of ten enough). 

Thus  each organism is on a journey. There is always some thing else  to be experienced within the physical hierarchy, and therefore there are always further mental elements yet to be experienced within the corresponding power set; always something that each organism has yet to experience and does not  yet {\it know}. Each is on a journey: but at any particular moment in time it has only every encountered a finite number of physical elements,  say $n$  of them, and thus its present set of mental elements is also finite: there would almost be $2^n$ of them.


\section{Summary: implications of the Dual Hierarchy model}
Why do we need a mental elements hierarchy? Why cannot we all be appreciating zombies, relying merely upon  the physical elements hierarchy and and learn to make it as complete as possible? We begin  a response by considering analogous constructs in efficient problem solving.  

The mental elements are equivalent to unobservable {\it latent variables} deployed  in multi-hypothesis design making, multimode dynamics,  and control problems. Without such latent modes, in order to switch from one behavioural mode to another, say from flight to fight, an organism  would need a huge amount of physical evidence. 

The {\it latency} concept is fundamental in hidden Markov models, for example. The existence and switching of the latent (hidden) states, would help achieve whole organism  behavioural changes far more quickly than bottom up reasoning. So in order to be effective and efficient even zombies or  computerised robot minds would be best to include a rather wide range of  latent states, operating over both small and large domains.  Indeed dual systems are very prevalent throughout both control (for example, of Hamiltonian systems and bang-bang control theory) and optimisation theory (such as in linear programming and so on). Often it is far more efficient to resolve issues within the dual system (in our case the dual hierarchy) to drive and focus the physical one.  We contend that the dual hierarchy system evolves to serve this purpose and is embedded within the architecture as opposed to being instantiated in processing work.


Our suggestion is the process of evolution would consign advantages to organisms capable of resolving problems within such a dual system approach, making up for incomplete information and resolving behavioural response rapidly and top down (in the mental) as the same time as bottom up (in the physical),  in perceiving what is physically happening. This last is really a {\it non-constructive argument for the existence of a dual hierarchy } (contrasting with the constructive basis given in the previous section).

An interesting difficulty is one of counting (of making lists), and possibly of a countable infinity. If we limit a mental element $Y$  to be  a finite set of justifying or associated elements, then we fail: a zombie or a robot could easily learn all of those elements  in $Y$ by rote. There would be  nothing that  ``Mary didn't know'' \cite{frank}. Yet if $ Y$ includes an uncountable number of elements then even a lifetime could never be enough to list or assemble them (there must always be element that ``Mary didn't know'').  Even so $Y$ exists within the mental hierarchy (involving a possible countable infinity of partially defining physical elements, which are available of course due to recursion). Only those mental states corresponding to a countable infinity of physical elements lie beyond the learning abilities of robots or zombies (or Mary). Such states must be experienced and perceived through an organism's own recursive processes and the mapping from the physical hierarchy  onto the mental hierarchy. 

Each individual may generate a larger and larger physical elements network over a lifetime, as a result of both perceiving more and more low domain (low level) objects and actions, as well as through the unconscious exercise of recursion creating relations between relations between relations between objects and so on. Then its mental elements hierarchy (potentially the power set of the set of physical elements) also grows with it - exponentially so.  It is also possible that the elements, and thus levels within the hierarchies, might become unlearned - perhaps as a result of cognitive decline or else a lack of continual usage. 

How many separate physical and mental elements might a human brain really hold at any one time?   While this is very very large, it must however remain finite. Thus, though the dual hierarchy model is infinite in concept, in practice every organism is on a expanding (or contracting) journey. At any moment there are always experiences that have not been felt and are unknown to each organism - though each has its own superset. The difference between a human brain and a zombie is that the brain has the potential to include any elements, as appropriate, in time. For the avoidance of doubt we do not propose that a brain could exploit the recursive machinery at its own convenience and apply it to  generate real elements on the fly  as desired: recall the recursion is instituted and hardwired with connections between SCDSGs.

A human brain presently restricted to a finite set of observations, stimuli or experiences,  must be unconscious of  those mental and physical elements of which it has no knowledge, that lie beyond it.  
A  human brain is continuously learning and experiencing, and so it has the potential at least in time to become more and more conscious, and to reach higher and higher domains. Consciousness is thus a relative concept and is best spoken of  as journey within the dual hierarchy rather than an achievable destination.

For those insisting there is still an ``explanatory gap'', we emphasise that for some feelings there can be no finite listing and thus no full learning of everything that is associated with that subjective experience. Each individual is working though a possibly longer and longer list of learnings about what such an experience involves (what it is like to see blue): but it will never be compete, even in a lifetime. Thus the ``explanatory gap'' does not refer to what is indicated here, in terms of a possible infinite structures of physical and mental elements  and what is {\it felt} in a subjective way. Instead it refers to what is achievable here through any (finite) learning and experience by an  individual compared with  the infinite {\it whole nine yards}, which none of us will every experience. 

It is therefore not for us, in proposing that this structure is capable of reflecting all facets of mental subjective feelings (even infinitely many), to show that the mental elements correspond  to subjective feelings; but it is for those opponents of the model to bring forward any one single  facet of any subjective experience  which could not be represented within the dual hierarchy, and thus experienced at some point by organisms that are learning and growing thus. 

\section*{Acknowledgements}
I am grateful for support from a number of EPSRC research grants: 
 EP/G065802/1, {\it The Digital Economy HORIZON Hub}; 
 EP/I017321/1, {\it MOLTEN: Mathematics Of Large Technological Evolving Networks};  
 EP/F033036/1, {\it Cognitive Systems Science (Bridging the Gaps)}; 
  EP/I016856/1,  {\it NeuroCloud};  and 
  EP/H024883/1, {\it Towards an integrated neural field computational model of the brain}. I am also pleased to acknowledge the advice and encouragement of my colleagues on those grants,  especially Doug Saddy and Des Higham, along with many helpful  challenges to my thinking from  Clive Bowman.

\end{document}